\documentclass{article}

\usepackage{microtype}
\usepackage{graphicx}
\usepackage{subfigure}
\usepackage{booktabs} 
\usepackage[hyphens]{url}
\usepackage{tikz} 
\usepackage{hyperref}
\usepackage[normalem]{ulem} 

\usepackage[accepted]{icml2019}

\icmltitlerunning{A Flexible Pipeline for Data-Driven Prediction of Tropical Cyclone Paths}

\begin{document}

\twocolumn[
\icmltitle{A Flexible Pipeline for Prediction of Tropical Cyclone Paths}



\icmlsetsymbol{equal}{*}

\begin{icmlauthorlist}
\icmlauthor{Niccol\`o Dalmasso}{equal,cmu}
\icmlauthor{Robin Dunn}{equal,cmu}
\icmlauthor{Benjamin LeRoy}{equal,cmu}
\icmlauthor{Chad Schafer}{cmu}
\end{icmlauthorlist}

\icmlaffiliation{cmu}{Department of Statistics \& Data Science, Carnegie Mellon University, Pittsburgh, PA, USA}

\icmlcorrespondingauthor{Benjamin LeRoy}{bpleroy@stat.cmu.edu}

\icmlkeywords{Weather Modeling, Prediction Bands, Spatiotemporal Data, Data Visualization} 

\vskip 0.3in
]



\printAffiliationsAndNotice{\icmlEqualContribution} 


\section{Introduction}

Hurricanes and, more generally, tropical cyclones (TCs) are rare, complex natural phenomena of both scientific and public interest. The importance of understanding TCs in a changing climate \cite{Knutson2010} has increased as recent TCs have had devastating impacts on human lives and communities \citep{ABC11florence, Americanredcrossmozambique}. Moreover, good prediction and understanding about the complex nature of TCs can mitigate some of these human and property losses. Though TCs have been studied from many different angles, more work is needed from a statistical approach of providing prediction regions \cite{camargo2016tropical}. The current state-of-the-art in TC prediction bands comes from the National Hurricane Center of the National Oceanographic and Atmospheric Administration (NOAA), whose proprietary model provides ``cones of uncertainty'' for TCs through an analysis of historical forecast errors \cite{nhc}. 

The contribution of this paper is twofold. We introduce a new pipeline that encourages transparent and adaptable prediction band development by streamlining cyclone track simulation and prediction band generation. We also provide updates to existing models and novel statistical methodologies in both areas of the pipeline, respectively. 


Our pipeline has many desirable properties. 
It provides an \textit{easy-to-use} framework to estimate prediction bands for new TCs. 
The pipeline consists of two separable components for TC track simulation and prediction band creation. 
Each individual part is designed to be \textit{interchangeable}; provided the input and output are the same, each part can easily be replaced by the user's favorite method. 
We provide \textit{purely data driven} TC simulation approaches, but one can easily pair these with other meteorological methods. This is especially important in a climate change scenario, where adapting to relatively quick changes in TC behavior is potentially key for damage mitigation.
Our current pipeline is implemented as an \texttt{R} package and is available on \texttt{Github}\footnote{\href{https://github.com/Mr8ND/TC-prediction-bands/}{github.com/Mr8ND/TC-prediction-bands/}}. Section \ref{sec:pipeline} details its structure and components.

We extend the current statistical methods for TC simulations, combining autoregressive path prediction models \cite{vickery2000simulation} and models of TC lifespan \cite{hall2007statistical}.
We propose novel statistically valid prediction bands by leveraging geometric properties and statistical notions on the depth and centrality of simulated curves. We benchmark against the prediction bands creation technique descripted in NOAA's state-of-the-art model \cite{nhc}.
We construct prediction band methods that provide coverage in either a pointwise or uniform manner, i.e., aiming to cover either the future measurements of the TC path or the full TC track, respectively. Section \ref{sec:methods} discusses these contributions in detail, highlights experimental results on held-out data, and provides empirical assessments of the prediction band methods' statistical properties.








\section{Pipeline Structure and Flow} \label{sec:pipeline}

\begin{figure*}[!ht]
    \centering
    \includegraphics[width = 1\linewidth]{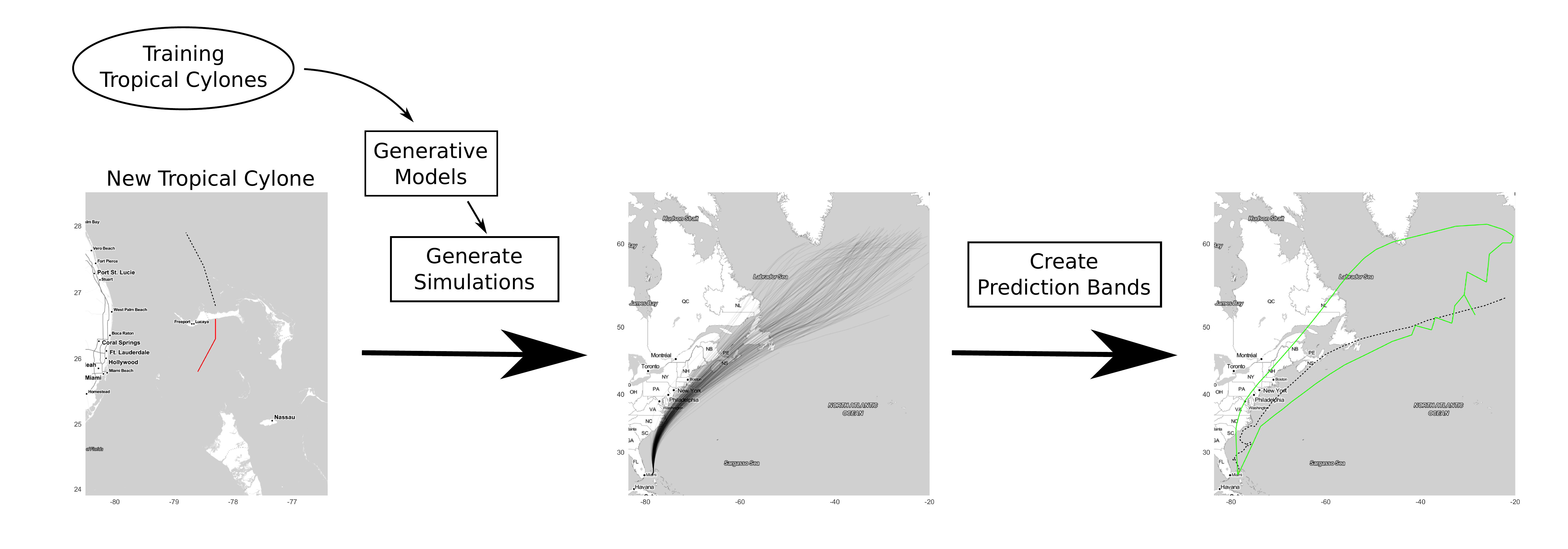}
    \vspace{-6mm}
    \caption{Pipeline structure schema. From left to right, when the start (6-12 hours) of a new tropical cyclone is observed, a pre-trained model generates track simulations, and the prediction band methods use the simulated ensemble to construct the prediction band.}
    \label{fig:pipeline}
\end{figure*}

Figure \ref{fig:pipeline} provides a schematic representation of our proposed pipeline.
The leftmost stage corresponds to the input, which is a series of spatiotemporal measurements of a TC track. 
In the current iteration we rely on the Atlantic Oceanographic and Meteorological Laboratory of NOAA, which publicly shares data on TC paths in the North Atlantic \citep{landsea2013atlantic, hurdat2}. Each path has information on time, latitude, and longitude, recorded every six hours until the storm dies.
For each new TC, the pipeline uses only a few initial points to simulate a series of tracks the TC could traverse. In this paper, we showcase the pipeline's potential to predict the full TC track, but the overall structure can also support forecasting shorter time frames, which is more realistic in terms of applications. 
When training the track simulation model, the training TC paths should be representative of the world region for which the pipeline will make predictions. For instance, TC track data over the Atlantic Ocean should not be used to train models over the Pacific Ocean.
Once these simulations are generated, the next stage uses this ensemble to produce a prediction band at a chosen statistical confidence level, usually between $90\%$ and $99\%$. The confidence refers to the statistical coverage of the prediction band, i.e., the rate at which the prediction band covers the true TC points or the full TC track. When considering different prediction band techniques, one could use a held-out set of TC tracks to assess the performance, taking into consideration the band coverage and size. 

It is important to reiterate that the pipeline is model-agnostic. The choices of statistical framework or machine learning models do not affect the structure or flow of the pipeline, provided the input and output of each stage is not changed. Similarly, a more granular or richer input will not modify the pipeline structure but will only affect the models in the simulation and band prediction stages.

\section{Current Methodology} \label{sec:methods}


\begin{table*}[!ht]
\centering
\begin{tabular}{p{1.5in}p{.95in}p{.95in}p{.95in}p{.95in}}
  \hline
\hline
Simulation Curve Type & $\delta$-Ball & Convex Hull & \hbox{Kernel Density} Estimate & Spherical  \\ 
  \hline
  AR \& Logistic     & 0.88 / 0.43 & 0.75 / 0.33 & 0.50 / 0.14 & 0.31 / 0.03   \\ 
  AR \& Kernel       & 0.88 / 0.44 & 0.86 / 0.37 & 0.55 / 0.19 & 0.31 / 0.04  \\ 
  Non-AR \& Logistic & 0.85 / 0.40 & 0.72 / 0.33 & 0.45 / 0.13 & 0.24 / 0.03  \\ 
  Non-AR \& Kernel   & 0.88 / 0.43 & 0.79 / 0.39 & 0.50 / 0.16 & 0.24 / 0.03  \\ 
   \hline
\end{tabular}
\caption{Median proportion of true test TC points captured by their prediction bands / proportion of test TCs with 100\% of points captured. Based on prediction bands attempting to obtain $90\%$ coverage.}
\label{tab:capture_rates}
\end{table*}

Our methodology for predicting TC paths involves two major modeling components: the simulation of potential paths and the construction of prediction bands from those paths.

To simulate TC paths, first we have developed models that make sequential predictions for the location of a TC. Our current approaches build upon foundational work by Vickery et al., who predicted change in TC bearing and speed from linear models with basin-specific coefficients \cite{vickery2000simulation}. By sequentially predicting changes in TC bearing and speed at each time step, these models propagate complete TC paths from the initial points. Whereas Vickery et al.\ include two lag terms in their bearing model and one lag term in their speed model, we have developed two sets of models. The non-autoregressive (non-AR) models predict change in bearing and change in speed at time step $i$ from latitude (at time step $i$), longitude, bearing (calculated between time steps $i-1$ and $i$), and speed (between time steps $i-1$ and $i$). The autoregressive (AR) models also include a lag term for change in bearing or change in speed, respectively. Conceptually, the AR models allow us to model climatological phenomena such as TCs continuing to gain/lose speed or continuing to turn in the same direction. Empirically, table \ref{tab:capture_rates} shows that the prediction bands constructed from AR simulations capture greater proportions of the test TC points and full paths.
To complete our generative models, we also develop models for TC lysis (death). Our first lysis model is a block-specific logistic regression. Earlier work on Gaussian weighting of historical TC lysis locations showed the promise of location-dependant lysis models \cite{hall2007statistical}. The logistic regression model predicts TC death at each point based on latitude, longitude, bearing, and speed. We also include a kernel density approach to lysis modeling, which draws TC lengths from a smoothed distribution of TC lifespans. In table \ref{tab:capture_rates}, we see that kernel density lysis generally yields slightly higher coverage rates on test TCs.

We train and evaluate our models on 1008 TCs from the NOAA database between June 1851 and November 2016 \citep{hurdat2}. We use 702 TCs to train the bearing, speed, and lysis models on 10-degree latitude-longitude squares. Then we generate potential paths from the initial observations of the 306 test TCs. To predict trajectories from a non-AR model, we initialize the path with the first two TC observations (6 hours), to observe speed and bearing. To predict trajectories from an AR model, we start with the first three TC observations (12 hours), to observe change in speed and change in bearing.
For each new/test TC we generate a set of curves (350 for current analysis) using a combination of AR or non-AR model and lysis method.
Upon simulating 350 potential TC paths for each test TC under each method, we construct prediction bands for the TC paths.  This is similar to the approach of the National Hurricane Center of NOAA, which predicts circular regions in which a TC center will fall with high probability, up to five days out from the point of prediction \citep{nhc}. In contrast, our approach aims to predict the location of the TC across its entire lifespan. 

To construct prediction bands, we have developed two approaches based on pointwise confidence intervals and two approaches based on uniform confidence structure. One pointwise approach constructs a kernel density from the set of all simulated points; the prediction band is the $100(1-\alpha)\%$ level set at a pre-specified coverage level. The second pointwise approach uses a depth metric proposed by Geenens and Nieto-Reyes \cite{geenens2017depth} to determine the most central simulated path with desirable structure for functionals \cite{nietoreyes2011}. At each time step $t$, this approach finds the smallest sphere, centered at the $t^{th}$ point of the most central path, that contains $100(1-\alpha)\%$ of the other simulated TCs' points at time $t$. We connect these spheres to form the prediction band. Out of the four approaches, this spherical ball method is closest to the approach of the National Hurricane Center of NOAA \cite{nhc}. The uniform approaches treat the entire simulated tracks as the objects of interest. One uniform approach creates a prediction band from the convex hull of the top $100(1-\alpha)\%$ of deepest curves. The second uniform prediction band contains a set of balls centered at all points in the top $100(1-\alpha)$\% of deepest curves. The balls have a common radius ($\delta$), which is the smallest radius such that all balls are connected to at least one other ball. We call this prediction band method the $\delta$-ball method. 

\begin{figure}
    \centering
	\input{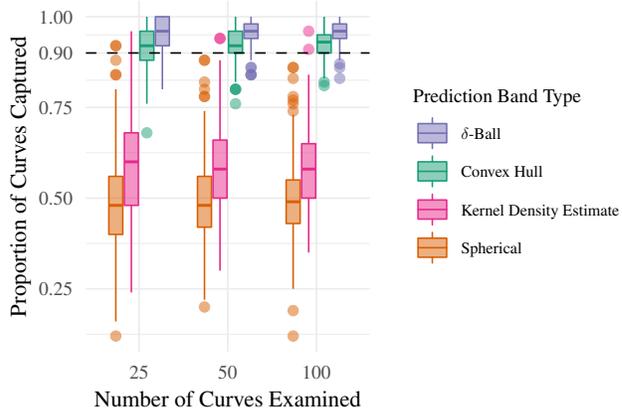}
	
	\vspace{-7mm}
	\caption{Uniform coverage of prediction bands, i.e. proportion of times the prediction band captures the whole TC track, generated from autoregressive and logistic-based lysis models. Pointwise prediction bands are intrinsically not well suited for this task.}
	\label{fig:coverage}
	\vspace{-3mm}
\end{figure}

In table \ref{tab:capture_rates}, we see that the $\delta$-ball prediction bands have both the highest pointwise and uniform TC coverage, across all simulation models at $\alpha = 0.10$. The difference between logistic and kernel lysis coverage is negligible for the $\delta$-ball TCs. Since the logistic approach incorporates location information and tends to yield PBs with smaller area, we recommend the $\delta$-ball approach with logistic lysis and AR simulation models. The $\delta$-ball prediction bands are also smaller than the convex hull prediction bands (the other proposed uniform prediction band method). One can notice that the $\delta$-ball prediction bands still achieve less than $90\%$ pointwise and uniform coverage rates. As a prediction band method, figure \ref{fig:coverage} shows that the $\delta$-ball method achieves proper uniform coverage under experimental simulations. Specifically, figure \ref{fig:coverage} shows the coverage results of  experiments in which we construct the prediction band-generating TCs and the ``true'' TC from the same methods. This suggests that the less than $90\%$ coverage rates may be due to a failure of the generative models to capture the true distribution of TC paths. Nevertheless, the $\delta$-ball prediction bands with AR simulation models and kernel lysis are able to capture a median of 88\% of test TC points, and they capture the full path for 44\% of the test TCs.


\section{Future Work and Extensions} \label{sec:extensions}
We anticipate the modeling components to be expanded to include more extensive and potentially nonparametric time series models. From a statistical perspective, the current methodologies fall within a parametric bootstrap framework, but Bayesian models such as multi-dimensional Gaussian processes could be implemented, given the pipeline's flexibility. Incorporating inputs from different scientific probes (e.g., satellite images, oceanic surface temperatures) and different climatological models could also be feasible, by establishing parallel threads within the same pipeline.

\textbf{Acknowledgments} Niccol\`o Dalmasso and Chad Schafer are partially supported by the National Science Foundation under Grant Nos.\ DMS1521786. Robin Dunn is supported by the National Science Foundation Graduate Research Fellowship Program under Grant Nos.\ DGE1252522 and DGE1745016.

\end{document}